\documentclass[superscriptaddress,aps,prX,twocolumn,showpacs,nofootinbib,longbibliography,notitlepage,floatfix]{revtex4-2}
\usepackage{etex}
\usepackage{amsmath,amssymb,amsthm}
\usepackage[colorlinks=true,citecolor=blue,urlcolor=blue]{hyperref}
\usepackage[pdftex]{graphicx}
\usepackage{times,txfonts}
\usepackage{braket}
\usepackage{color}
\usepackage{natbib}
\usepackage{amsmath,blkarray}
\usepackage{mathtools}
\usepackage{ulem}
\usepackage{latexsym}
\usepackage{tabularx, booktabs}
\usepackage{graphics,epstopdf}
\usepackage{graphicx}
\usepackage{float}
\usepackage{amsfonts}
\usepackage{tikz}
\usetikzlibrary{quantikz}
\usepackage{color,soul}

\newcommand{\be}{\begin{equation}}
\newcommand{\ee}{\end{equation}}
\newcommand{\ba}{\begin{eqnarray}}
\newcommand{\ea}{\end{eqnarray}}

\newcommand{\tr}{\operatorname{Tr}}

\newcommand{\etal}{{\it{et al. }}}

\usepackage{multirow}
\usepackage{appendix}
\usepackage{url}

\begin{document}

\title{Portfolio Optimization with Digitized-Counterdiabatic Quantum Algorithms} 

\author{N. N. Hegade}
\thanks{Co-first authors.}
\affiliation{International Center of Quantum Artificial Intelligence for Science and Technology~(QuArtist) \\ and Physics Department, Shanghai University, 200444 Shanghai, China}

\author{P. Chandarana}
\thanks{Co-first authors.}
\affiliation{Department of Physical Chemistry, University of the Basque Country UPV/EHU, Apartado 644, 48080 Bilbao, Spain}

\author{K. Paul}
\affiliation{International Center of Quantum Artificial Intelligence for Science and Technology~(QuArtist) \\ and Physics Department, Shanghai University, 200444 Shanghai, China}

\author{X. Chen}
\affiliation{Department of Physical Chemistry, University of the Basque Country UPV/EHU, Apartado 644, 48080 Bilbao, Spain}

\author{F. Albarr\'an-Arriagada}
\email{pancho.albarran@gmail.com}
\affiliation{International Center of Quantum Artificial Intelligence for Science and Technology~(QuArtist) \\ and Physics Department, Shanghai University, 200444 Shanghai, China}

\author{E. Solano}
\email{enr.solano@gmail.com}
\affiliation{International Center of Quantum Artificial Intelligence for Science and Technology~(QuArtist) \\ and Physics Department, Shanghai University, 200444 Shanghai, China}
\affiliation{IKERBASQUE, Basque Foundation for Science, Plaza Euskadi 5, 48009 Bilbao, Spain}
\affiliation{Kipu Quantum, Kurwenalstrasse 1, 80804 Munich, Germany}

\date{\today}

\begin{abstract}
We consider digitized-counterdiabatic quantum computing as an advanced paradigm to approach quantum advantage for industrial applications in the NISQ era. We apply this concept to investigate a discrete mean-variance portfolio optimization problem, showing its usefulness in a key finance application. Our analysis shows a drastic improvement in the success probabilities of the resulting digital quantum algorithm when approximate counterdiabatic techniques are introduced. Along these lines, we discuss the enhanced performance of our methods over variational quantum algorithms like QAOA and DC-QAOA.

\end{abstract}

\maketitle

\section{\label{intro} Introduction}
Optimization problems have been of significant interest due to their fundamental applications in many fields such as logistics, medicine, finance, among others. Nevertheless, due to their computational complexity, they cannot be solved efficiently using classical computers for industrial purposes. It is believed that a quantum computer might surpass the capabilities of a classical one. Due to the experimental developments during the last years, it might become useful for commercial purposes~\cite{Mohseni2017Nature, Bova2021EPJQT}. This potential breakthrough has boosted proposals of several algorithms in different areas, such as differential equations, linear algebra, and optimization problems~\cite{Montanaro2016NPJ}, which have been implemented in small quantum computers as proof of principle. However, the applicability of these algorithms in the current noisy intermediate-scale quantum (NISQ) devices~\cite{preskill} is still under investigation. This is because of the difficulty to implement scalable error-correction protocols with the current and near-future devices, a bottle neck for fault-tolerant quantum computing. 

In recent years, the use of adiabatic quantum optimization (AQO) algorithms to solve optimization problems has received interest due to its experimentally feasible implementation~\cite{qaa1,qaa2,qaa3,qaa4,qaa5}. These algorithms solve optimization problems by codifying them in the ground state of a Hamiltonian and accessing it via adiabatic evolution from another Hamiltonian, whose ground state is trivial to prepare. Current technology allows the implementation of incoherent adiabatic quantum computers or quantum annealers with thousands of qubits. Nevertheless, due to the considerable time involved in an adiabatic evolution, such devices have various limitations like noise and limited qubit connectivity. To overcome these difficulties, the use of digitized-adiabatic quantum computing (DAdQC)~\cite{Barends2016} methods has been developed and implemented. These techniques are similar to AQO, but they digitized the evolution to implement the corresponding algorithms in a gate-based quantum computer. The convenience of digitization is that any arbitrary interactions can be included in the target Hamiltonian, providing more flexibility in choosing the optimization problem. However, it requires a large number of gates, which reduces the fidelity of the algorithms and makes them still impractical to approach quantum advantage in NISQ devices without error correction.

Other types of algorithms to solve optimization problems are the hybrid quantum-classical gate-based algorithms, like quantum approximate optimization algorithm (QAOA)~\cite{farhi2014quantum}. In QAOA, a sequence of two evolutions is applied iteratively to an initial state. The evolution times are considered as free parameters to be optimized to minimize the cost function that encodes the solution to the problem. These evolutions are governed by a mixing Hamiltonian. Typically, a Pauli-X operation is applied to all the qubits and a problem Hamiltonian that codifies the cost function. Finally, the elapsed time of each evolution is optimized by a classical algorithm. Although QAOA is simple, it presents some major problems like the number of algorithmic layers required to optimize the cost function. This issue, in general, is hard to solve for many-body Hamiltonians, limiting the possible scalability in current NISQ devices. Additionally, classical optimization does not guarantee the reach of the global minimum, as it can get stuck into local minima, or barren plateaus may appear~\cite{barr1,barr2,barr3}.

To overcome these limitations, shortcuts to adiabaticity (STA) methods were proposed. These are known to improve the adiabatic processes by circumventing the need of slow driving~\cite{staa1}. STA includes methods like fast-forward~\cite{stab2,stac3}, invariant based inverse engineering~\cite{stad4,stae5}, and counterdiabatic (CD) driving~\cite{staf6,sta8,prielinger2021two,del2013shortcuts}. Among these, CD driving, i.e., the addition of an extra term that suppresses the non-adiabatic transitions, has already shown improvements in adiabatic quantum computing~\cite{naren2020,hegade2021digitized} and quantum annealing~\cite{ann1,ann2,ann3}.  

Recently, Hegade \etal showed the advantage of CD driving in DAdQC methods~\cite{naren2020}, which have shown interesting improvements in many-body ground state preparations and adiabatic quantum factorization problems \cite{hegade2021digitized}. It was also studied that the inclusion of these approximate counterdiabatic terms could enhance the performance of QAOA while solving combinatorial optimization problems and state preparation of many-body ground states \cite{reinfo, chandarana2021,wurtz2021counterdiabaticity}. This article considers the advantages of digitized-counterdiabatic quantum computing (DCQC) and digitized-counterdiabatic quantum approximate optimization algorithms (DC-QAOA) in financial applications. In particular, we investigate the Markowitz portfolio optimization problem~\cite{markowitz}, a common problem used intensively by financial asset managers. This problem deals with how to optimize the weights of the assets in a portfolio to give out returns based on the requirements of the asset manager, as is the case of maximum return and minimum risk. We select a large number of data instances and solve the problem using the DAdQC combined with CD-driving protocols. By considering variational minimization techniques, we obtain the approximate CD terms, getting a drastic enhancement in the success probability for most of selected instances. We also show that inclusion of these CD terms in hybrid quantum-classical algorithms like QAOA, namely DC-QAOA, results in improvement of success probabilities. This article demonstrates the relevance of CD driving, its limitations, and applicability for finance industry problems in the NISQ era.   

The manuscript is arranged as follows. In the following section Sec.~\ref{sec2}, we present the theory of the Markowitz portfolio optimization problem and how to encode it as a quadratic unconstrained binary optimization problem, which can later map to finding the ground state of an Ising Hamiltonian. In Sec.~\ref{sec3}, we formulate the digitized-counterdiabatic techniques to enhance the performance of adiabatic quantum algorithms for obtaining the approximate ground states in a finite time. We consider local CD driving and approximate CD terms and show their performance for solving different instances of the portfolio optimization problem. Sec.~\ref{sec4} considers the hybrid quantum-classical algorithms like QAOA and DC-QAOA to tackle the same problem and compare their performance. Finally, In Sec.~\ref{sec5}, we discuss our results and conclude.     

\section{PRELIMINARIES}\label{sec2}

Suppose an asset manager is willing to invest a budget $b$ in a given portfolio with $n$ number of assets. Markowitz portfolio optimization~\cite{markowitz} answers the question: How to distribute the given budget $b$ into $n$ assets such that the asset manager would receive maximum returns at minimum risk. For example, the asset manager could invest evenly in all the assets, but that may not be the best choice. Given the expected returns of each asset $x_i$ and the risk associated with the assets, applying Markowitz portfolio optimization, we can predict the distribution of the budget $b$ to get the maximum returns out of the portfolio. Expected returns can be estimated from the preceding market return data, and the risk is assessed via a covariance matrix. As the name suggests, a covariance matrix shows the covariance of different stocks in the portfolio. Thus, these types of problems are classified as mean-variance portfolio optimization problems.

The portfolio optimization problem has two components, the cost function and the constraints. The cost function includes quantities such as the average returns, the variance of the portfolio, and others that need to be maximized or minimized based upon the need of the asset manager, while the constraints can include budget constraints or transition costs or market inflation. Based on the constraints, the portfolio optimization problems can be broadly divided into three types. The first type of optimization problem is the unconstrained portfolio optimization, where the constraints are added as the penalty terms in the cost function with the help of Lagrange multipliers. The second type is when the constraints can be represented as inequalities, and the third type is when the constraints are the integer constraints, the so-called mixed-integer problem. Examples and computational complexities of all these problems are discussed in Ref.~\cite{bouland2020prospects}. Different types of these portfolio optimization problems have been solved recently using quantum computing~\cite{hodson2019portfolio,bench2021,solv2016,rev2019,marzec2016portfolio,rebentrost2018quantum,orus2019quantum,mugel2020dynamic}.

As far as the variables are concerned, portfolio optimization can be solved as a continuous-variable and discrete-variable problem. Discrete mean-variance optimization proves to be a beneficial method to optimize the given portfolio in cases where the assets are traded in lots, which are integer multiples of a base size of assets traded. Thus, the asset manager will only be interested in the number of lots which makes the problem discrete. In consequence, we will investigate here an unconstrained single-period discrete mean-variance portfolio optimization problem using the digitized-counterdiabatic quantum computing paradigm.  

Our task is to distribute $b$ budget into $n$ assets with mean returns $m_i$ and the co-variance matrix $\rho_{ij}$ to maximize returns at a minimum variance. The problem can be formulated as, 
\begin{equation}\label{cost1}
\max _{x} \sum_{i=1}^{n} \theta_1m_{i} x_{i} - \theta_2\rho_{ij}x_i x_j - \theta_3 ( G_f b x_{i}-b)^2 ,
\end{equation}
where $x_i$ are the assets represented by integers. The first term of Eq.~\eqref{cost1} indicates the expected returns, where $m_i$ are the daily return data, and the second term shows the variance of the portfolio, where $\rho_{ij}$ show the variance of $j$th asset with $i$. The market data ($m_i$ and $\rho_{ij}$) is easily available and can be generated in various ways~\cite{load1,loader2}. The third term shows the constraint applied to the budget. Most of the times, asset managers have a certain budget to follow, so this term penalizes the solutions that do not satisfy the budget criteria. The term $G_f$ in Eq.~\eqref{cost1} shows the granularity function which is given by  $G_f= 1/2^{(g-1)}$, where $g$ is the number of slices of the budget. This implies that the fraction of the budget that can be chosen is $G_f b x_i$. To explain this, assume that we set $g=2$, so the budget is cut into two slices giving $G_f=1/2$. Therefore, we have $b/2$ weight of the budget that can be invested in required assets. In other words, granularity function decides how small a fraction of $b$ you can invest in a particular stock.  Thus, with increasing the value of slice $g$, we get the freedom to invest smaller fractions of the budget which will give us more precision while deciding the optimal results. $\theta_1$, $\theta_2$ and $\theta_3$ are the Lagrange multipliers and their role is to adjust weights of different terms to adjust constraints according to the manager's requirements. For instance, if the manager is less concerned about the risk and more interested about the returns, we can set $\theta_1$ higher than $\theta_2$ such that the expected return term is given more weight.

Once these terms are determined, we can convert Eq.~\eqref{cost1} into an Ising Hamiltonian. To do so, we start with transforming Eq.~\eqref{cost1} into a quadratic unconstrained binary optimization problem (QUBO) by encoding assets $x_i$ integers into an $g$ bit binary number. Apart from binary encoding, various ways of encoding like unary, sequential, and partition have been proposed~\cite{solv2016}. Hence, $x_i$'s are given by
\begin{equation}
    x_i = \sum_{k=1}^{g} 2^{k-1} z_{u(i,k)} \, ,
    \label{ztox}
\end{equation}
where $u(i,k)=(i-1)g+k$ and $z\in\{0,1\}$. This problem can be transformed into a QUBO problem by substituting $x_i$'s from Eq.~\eqref{ztox} into Eq.~\eqref{cost1}. After making some rearrangements and transformation $z_i=\frac{1}{2} (1 + \sigma^z_i)$, we get an Ising Hamiltonian given by
\begin{equation}
H_p=\sum_{i} h_{i}\sigma^z_{i} +\sum_{i,j} J_{i j} \sigma^z_{i} \sigma^z_{j} +\beta \, ,
\label{ising}
\end{equation}
where $ J_{i j}$ shows the interaction coefficient between the spins $i$ and $j$, while $h_i$ shows the coefficient associated with the local field along the $z$-axis. Explicit forms of these coefficients are given by
\begin{equation}
 J_{i j} =\frac{1}{4} (\theta_{2} b^{2} G_f^{2}+\theta_{3} \rho_{i, j}) \, ,
\end{equation}
\begin{equation}
h_{i} =\frac{(-\theta_{1} m_{i}-2 \theta_{2} b^{2} G_f)}{2}+\sum_{j} J_{i, j} \, ,
\end{equation}
with a constant $\beta$ that can be ignored. In this manner, we have transformed Eq.~\eqref{cost1} into an Ising Hamiltonian such that its ground state encodes the solution to the optimization problem. In the following sections we show application of digitized-counterdiabatic quantum computing methods to find this ground state.

\begin{figure*}
    \centering
    \includegraphics[width=\linewidth]{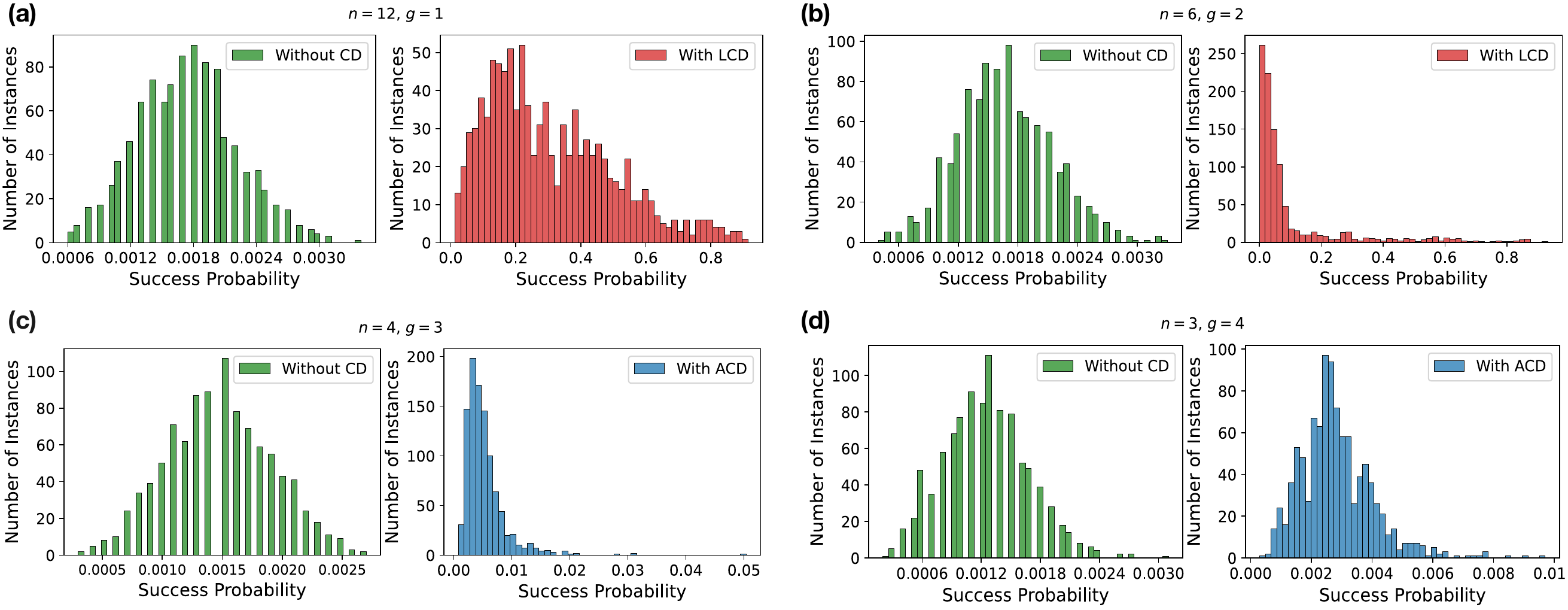}
    \caption{Histogram depicts the ground state success probability for 1000 randomly chosen instances of the portfolio optimization problem with and without CD driving. We considered 12 qubit systems with different numbers of assets ($n$) and partitions ($g$). When the number of partitions is smaller, the LCD in Eq.~\eqref{LCD} gives a better enhancement. Whereas, when $g$ is larger, ACD from Eq.~\eqref{ACD} shows greater improvement.  The simulation parameters are, total evolution time $T=1$, and the step size $\Delta t=0.05$.}
    \label{fig1}
\end{figure*}

\section{Digitized-Counterdiabatic Quantum Computing}\label{sec3}
To find the ground state of the Hamiltonian in Eq.~\eqref{ising}, we follow the adiabatic theorem by starting with an initial Hamiltonian whose ground state can be easily prepared, and slowly turn on the problem Hamiltonian. The corresponding total Hamiltonian is given by
\begin{equation}
    H_{ad}(\lambda) = (1-\lambda) H_i + \lambda H_p,
\end{equation}
where $\lambda$ is a time dependent scheduling function. We choose the initial Hamiltonian as $H_i = -\sum_i^N h^x \sigma_i^x$, whose ground state $\ket{\psi(0)} = 2^{-N/2}(\ket{0} + \ket{1})^{\otimes N}$ can be easily prepared. If the evolution is slow enough, the adiabatic theorem guarantees that the final state will have a large overlap with the ground state of the problem Hamiltonian. In principle, to obtain the ground state with a higher success probability, one has to consider the total evolution time $T$ much larger than the minimum energy gap $\Delta_{min}$ between the ground state and the first excited state. However, in practice, one can not rely on the adiabatic evolution due to limited coherence time, and device noise. One has to consider a short time evolution that will lead to non-adiabatic transitions between the eigenstates. In order to suppress these transitions, a technique called shortcuts to adiabaticity was developed \cite{torrontegui2013shortcuts,sta1}. The idea is to introduce an additional term called counterdiabatic-driving term (CD term) so that the excitations due to the finite time evolution will be compensated, and the resulting evolution will be quasi adiabatic. The modified Hamiltonian by adding the CD term takes the form
\begin{equation}
    H(t) = H_{ad} + \dot{\lambda} A_{\lambda} , 
\end{equation}
where $A_\lambda$ is called adiabatic gauge potential \cite{SelsE3909}, which satisfies the condition $\left[i \partial_{\lambda} H_{ad}-\left[{A}_{\lambda}, H_{ad}\right], H_{ad}\right]=0$. In principle, one can evolve the system very fast without any excitation by including the CD term. However, obtaining the exact gauge potential for a many-body system is a difficult task. Also, the operator form of the CD term for a many-body system with $N$ interacting spins generally contains non-local N-body interaction terms, which makes it experimentally challenging to realize. To overcome this challenge a variational method was proposed to obtain an approximate CD term \cite{SelsE3909}. And, many recent work shows the advantage of using variationally calculated local CD terms for different applications \cite{prielinger2021two,hartmann2020multi,iram2021controlling,hartmann2020multi}. The main advantage of this method is that, the approximate local CD terms can be easily implemented in lab and its calculation does not require any knowledge of the instantaneous eigenstates. This feature makes it suitable for adiabatic quantum computation.  

For the Hamiltonian \eqref{ising}, we choose the local CD term of the form $\Tilde{A}_{\lambda}=\sum_{i}^N \alpha_{i}(t) \sigma_{i}^{y}$. Here, the CD coefficient $\alpha_{i}(t)$ is obtained by minimizing the action $S =\tr[G_{\lambda}^2]$, where $G_{\lambda} = \partial_{\lambda} H_{ad} + i [\Tilde{A}_\lambda, H_{ad}]$,
\begin{equation}\label{LCD}
    \alpha_i(t) = -\frac{h^x h^z_i}{2 \left({h^x}^2 [\lambda -1]^2+\lambda^2 \left[{h^z_i}^2+ \sum_{i\neq j}J_{ij}^2\right]\right)}.
\end{equation}

The CD term should vanish at the beginning and end of the protocol, for that we consider the scheduling function as $\lambda(t)=\sin ^{2}\left[\frac{\pi}{2} \sin ^{2}\left(\frac{\pi t}{2 T}\right)\right]$, also $\dot{\lambda}$, and $\ddot{\lambda}$ vanishes at $t=0$ and $t=T$. The time evolution including the CD term is given by
\begin{equation}
\ket{\psi(T)} = \mathcal{T} e^{-i \int_{0}^{T} H(t) d t}\ket{\psi(0)}=U(0, T) \ket{\psi(0)} .
\end{equation}
For the gate-model implementation of the evolution, we write the total Hamiltonian as sum of $2$-local terms, i.e. $H(t) = \sum_j c_j(t) H_j(t)$. We discretize the total time into $M$ parts with step size $\Delta t = T/M$. Using Trotter-Suzuki formula, we approximate the time evolution operator as
\begin{equation}
    U_{\mathrm{dig}}(0, T) = \prod_{k=1}^{M} \prod_{j} \exp \left\{-i  \Delta t  c_j(k \Delta t) H_{j}\right\}.
    \label{evolution}
\end{equation} 

In this work, we only considered first-order trotterization, which leads to an error of the order $\mathcal{O}(\Delta t ^2)$. Using single-qubit and two-qubit gates, each matrix exponential term in Eq.~\eqref{evolution} can be easily implemented. Since the Hamiltonian involves all-to-all connection between the qubits, extra swap gates might be needed on a device with only nearest neighbour interactions. We consider a wide range of portfolios where the data ($m_i$ and $\rho_{ij}$) is randomly generated such that it mimics the real world trends. $\theta_1=0.3$, $\theta_2=0.5$ and $\theta_3=0.2$ are selected so that the budget constraint is given greater importance. We run 1000 different instances and system sizes upto $N=14$ qubits with different number of stocks $n$ and slicing $g$.

Since current NISQ devices can only implement circuits of limited depth, we consider a short-time evolution and compare the final success probability with and without including the CD term. In Fig.~\ref{fig1}, the ground state success probability for 1000 randomly chosen instances of the portfolio optimization problem is depicted for a system size $N=12$. In Fig.~\ref{fig1}a, the number of assets and the slices are chosen as $n=12$, $g=1$, respectively. And, in Fig.~\ref{fig1}b, $n=6$, $g=2$. We fix the total evolution time $T=1$, and step size $\Delta t = 0.05$ with total 20 Trotter steps. In both cases, the local CD (LCD) term $H_{LCD} = \dot{\lambda} \sum_{i}^N \alpha_{i}(t) \sigma_{i}^{y}$ is considered.

We observe that when the number of slices is less, the LCD terms drastically improves the success probability for most of the instances. However, as $g$ becomes larger, the performance of LCD terms will be decreased. Previously, it was shown that the density of states close to the ground state would also become larger with the increasing number of slices, leading to an increase in the probability of transition to the higher excited states \cite{bench2021}. In order to suppress these transitions, we considered higher-order CD terms obtained by the nested commutator (NC) ansatz \cite{PhysRevLett.123.090602},
\begin{equation}
    A_{\lambda}^{(l)} = i \sum_{k = 1}^l \alpha_k(t) \underbrace{[H_{ad},[H_{ad},......[H_{ad},}_{2k-1}\partial_{\lambda} H_{ad}]]] .
    \label{gauge}
\end{equation}
Here, $l$ corresponds to the expansion order, and when $l \to \infty$ we will get the exact gauge potential. 
By considering only the first-order expansion ($l=1$), we obtain the approximate CD (ACD) term for the problem Hamiltonian in Eq.~\eqref{ising} as

\begin{equation}\label{ACD}
    H_{ACD} =2 \dot{\lambda} h^x \alpha_1(t) \left[ \sum_i^N h_i^z {\sigma_i^y} + \sum_{i,j} J_{ij} (\sigma_i^y\sigma_j^z + \sigma_i^z \sigma_j^y)\right],
\end{equation}
where $\alpha_1(t)$ is obtained as before by minimization of the action $S =\tr[G_{\lambda}^2]$. The general expression for $\alpha_1(t)$ for an Ising spin-glass Hamiltonian is given in \cite{spinglass}. In Figs.~\ref{fig1}c and \ref{fig1}d, we considered $g=3$, and $g=4$, respectively, and compare the probability of success for different instances by including ACD and without including the CD term.
 
\begin{figure}
    \centering
    \includegraphics[width=\linewidth]{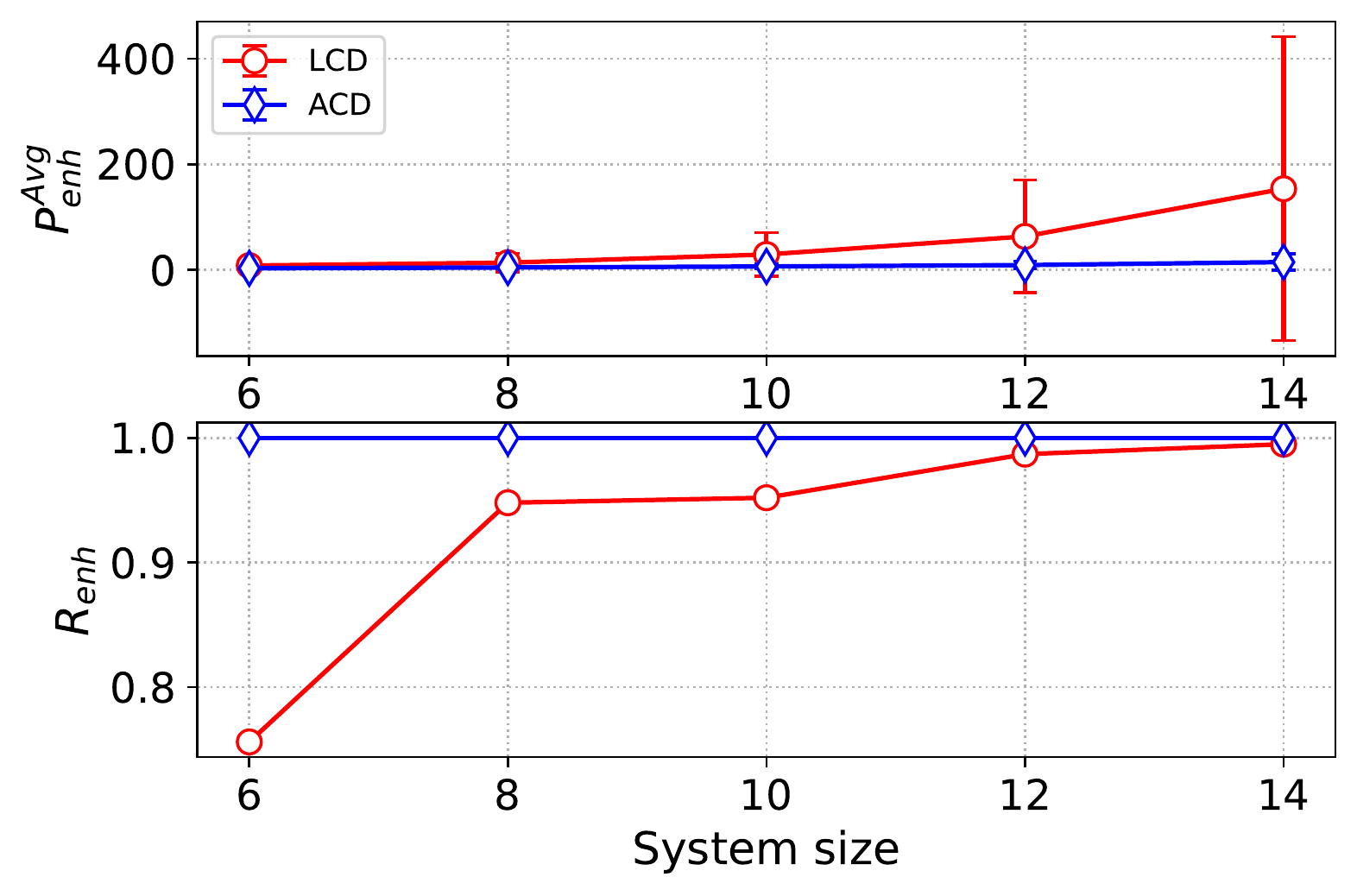}
    \caption{The upper plot depicts the average success probability enhancement (Eq.~\eqref{prob_enh}), and the lower plot represents the enhancement ratio (Eq.~\eqref{enh_ratio}) for various system sizes with LCD and ACD. The error bars represent the standard deviation. Due to the huge variations in the success probability enhancement for different instances using LCD, the error bars have larger values.}
    \label{fig2}
\end{figure}

\begin{figure}
    \centering
    \includegraphics[width=\linewidth]{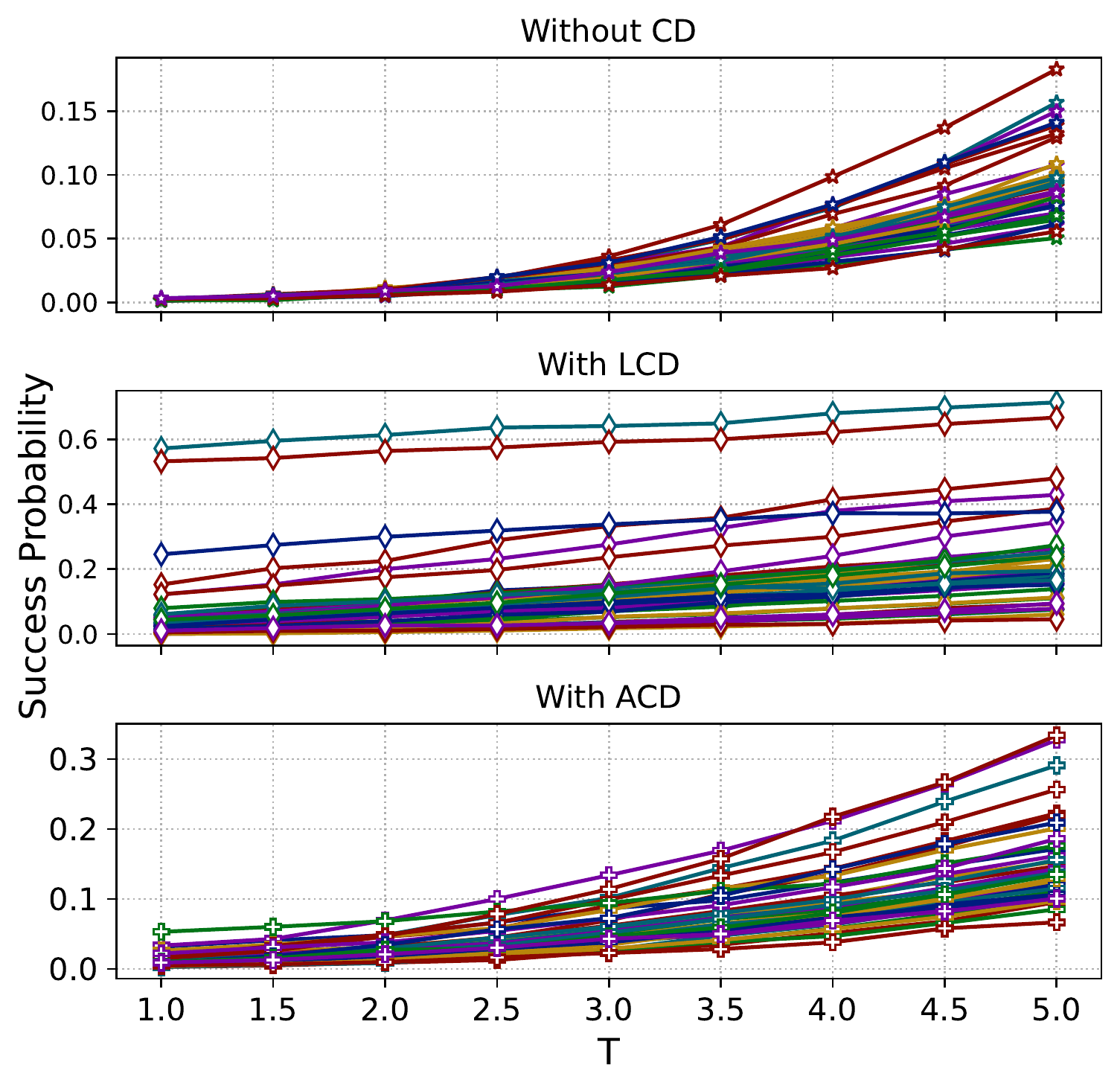}
    \caption{Success probability as a function of total evolution time $T$ for randomly chosen 40 instances is depicted. Here we considered $n=6$ assets and $g=2$ partitions resulting in system size $N=12$. A substantial improvement in success probability is obtained for most of the instances by including the local CD term $H_{LCD}$, and the approximate CD term $H_{ACD}$. Here we considered the Trotter step size as $\Delta t=0.1$.}
    \label{fig3}
\end{figure}

To show the enhancement that results from applying the CD term, we define a metric called enhancement ratio,
\begin{equation}
    R_{e n h}=\frac{I}{I^0} ,
    \label{enh_ratio}
\end{equation}
where $I^0$ is the total number of instances considered, and $I$ denotes the number of instances with enhanced success probability by including the CD term. 
The enhancement that results from applying the CD term is quantified by the success probability enhancement,
\begin{equation}
    P_{e n h}=\frac{P}{P^{0}} \, .
    \label{prob_enh}
\end{equation}
Here, $P$ is the ground state success probability by including the CD term, and $P^{0}$ is the success probability obtained by naive evolution without the CD term. Fig.~\ref{fig2} depicts the average success probability enhancement and enhancement ratio for various system sizes by including the CD driving. We observed that for ACD the enhancement ratio is ~1, indicating that it is always advantageous to include the approximate CD terms obtained from the first order NC ansatz with fixed $g=2$. The simulation result indicates that the enhancement with the CD term will increase with the system size. In Fig.~\ref{fig3}, the ground state success probability as a function of total evolution time $T$ is shown. The result shows that a huge enhancement can be obtained for both LCD and ACD. 

\section{Digitized-counterdiabatic quantum approximate optimization algorithm (DC-QAOA)}\label{sec4}
Hybrid quantum-classical optimization algorithms like QAOA~\cite{farhi2014quantum} have been of interest due to their applicability in the noisy intermediate-scale quantum (NISQ) era. These algorithms fall under the class of variational quantum algorithms, where classical optimization routines are employed to find suitable parameters $(\gamma,\beta)$ that optimize a cost function $C$. In our case, we have set $C =\bra{\psi_f(\gamma,\beta)}H_p\ket{\psi_f(\gamma,\beta)}$ so that $C$ shows the expectation value of the problem Hamiltonian $H_p$  and $\ket{\psi_f(\gamma,\beta)}$ is given by,
\begin{equation}
\ket{\psi_f(\gamma,\beta)} =  U( \gamma, \beta)\ket{\psi_i}
\end{equation}
where $\ket{\psi_i}= \ket{+}^{\otimes N}$ and,
\begin{equation}
    U(\gamma, \beta) = U_m{(\beta_p)}U_p{(\gamma_p)}U_m{(\beta_{p-1})}U_p{(\gamma_{p-1})} \dots U_m{(\beta_1)}U_p{(\gamma_1)}.
\end{equation}
Here, $p$ shows the number of layers, $U_m(\beta)$ corresponds to the mixer term, and $U_p(\gamma)$ corresponds to the problem Hamiltonian term. Over the years, many modifications have been reported in the standard QAOA~\cite{adapt,alter2019,daqaoa}. Among them, the addition of terms using counterdiabatic (CD) driving has shown significant improvements in finding ground states of many-body Hamiltonians~\cite{chandarana2021,reinfo,wurtz2021counterdiabaticity}. One of the algorithms following the same principle is the digitized-counterdiabatic quantum approximate optimization algorithm (DC-QAOA). In DC-QAOA, counterdiabatic (CD) driving is utilized to introduce an additional unitary $U_D(\alpha)$, known as the CD term. The form of CD term is chosen from the operator pool $A_{\lambda}^{(l)}$ obtained after performing the nested commutator (NC) method as mentioned in Sec.~\ref{sec3}. In this section, we perform QAOA and DC-QAOA for several instances of a small portfolio and compare the obtained success probabilities $(P_s)$ to demonstrate the advantage of CD driving over naive QAOA.

\begin{figure}
    \centering
    \includegraphics[width=1\linewidth]{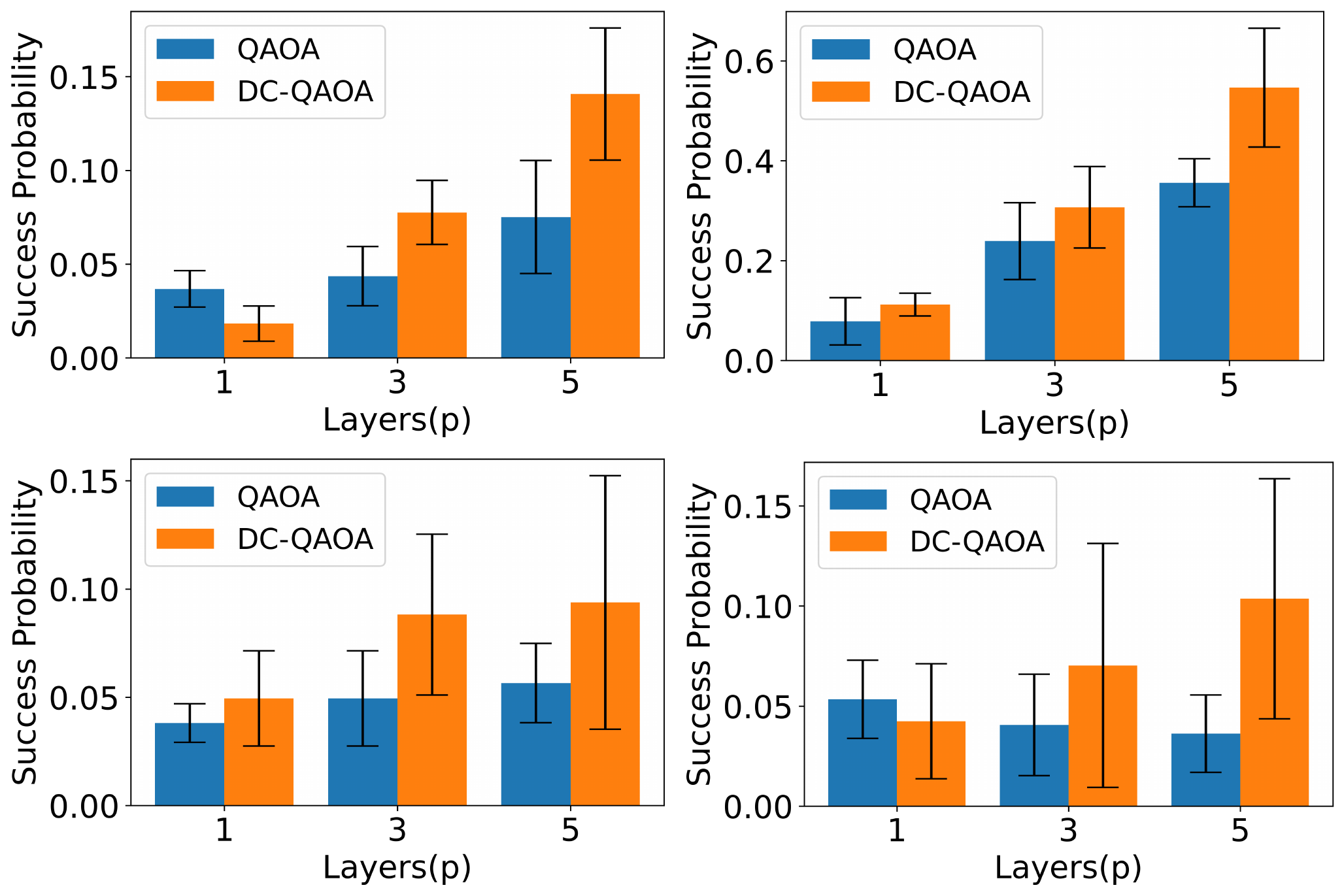}
    \caption{Success probability as a function of number of layers($p$) for $n=3$ and $g=2$ comparing performances of DC-QAOA and QAOA. Results shown are the average of top 10 out of 20 random patrameter initialization. Blue and orange bars show results of QAOA and DC-QAOA respectively. Error bars represent standard deviation. }
    \label{fig4}
\end{figure}

In QAOA, $p$-layers of mixer term $U_b(\beta)=   e^{-i\beta \sum_i \sigma_i^x}$ and  Hamiltonian term $U_p(\gamma)=e^{-i\gamma H_p}$, where $H_p$ is given by Eq.~\eqref{ising}, are applied alternatively to the initial state $\ket{\psi_i}$. In DC-QAOA, CD term $U_D(\alpha)= e^{-i\alpha \sum_i h_i \sigma_i^y}$ is added along with $U_b(\beta)$ and $U_p(\gamma)$, where $\sigma^y$ is a local first-order term chosen from an operator pool $A^{(2)}_{\lambda} = \{\sigma^y, \sigma^z \sigma^y, \sigma^y \sigma^z,  \sigma^x \sigma^y,  \sigma^y  \sigma^x\}$ which is obtained from Eq.~\eqref{gauge}. Parameters $(\gamma,\beta,\alpha)$ are updated using a stochastic gradient descent based classical optimizer called Adagrad optimizer~\cite{grad} with step size = 0.1. We select a system of size $N=6$, with $n=3$ and $g=2$ and perform the optimization for layers $p=1,3,5$. 

Fig.~\ref{fig4} shows the average success probabilities ($P_s$) as a function of number of layers ($p$) for four instances. Results show the average of best 10 runs chosen out of 20 runs where for each run, initial parameters $(\gamma,\beta,\alpha)$  are chosen randomly. We observe that by using DC-QAOA,  improvement in the $P_s$ values is achieved for all the instances. Also, DC-QAOA outperforms QAOA for every instance we study. The error bars which show the standard deviation associated with DC-QAOA get larger as we go to higher layered circuits. This behavior is observed because the choice of initial parameters becomes an important factor in getting high $P_s$ values. With a larger system size, the Hilbert space becomes huge and the energy landscape becomes complex therefore finding the global minima becomes a challenging task for the classical optimizer. Having said that, if the parameters are initialized near the global minimum of the energy landscape, the algorithm will give higher success probabilities ($P_s$). Hence, when the parameters are initialized randomly, a huge variation in $P_s$ can be observed because of the presence of many local minima, making it difficult to reach the global minima. Thus, this spectrum of $P_s$ values shows high standard deviation as evident from Fig.~\ref{fig4}. Higher $P_s$ values can be attributed to the fact that the expressibility of the DC-QAOA ansatz is better than the naive QAOA. We can also see that while most of the instances show success probabilities around $P_s=0.10$, there is an instance where $P_s=0.60$ can be achieved. This indicates that these algorithms can be favorable for tackling financial problems.

\section{Conclusion}\label{sec5}
We studied the financial portfolio optimization problem using recently proposed digitized-counterdiabatic quantum algorithms. We computed the approximate CD terms that can be easily implemented on any current gate-based quantum computer. We compared the ground state success probability for the evolution with and without including the CD terms, by fixing the total evolution time or the number of Trotter steps. We considered many instances of the portfolio optimization problem with randomly generated data. The results indicate that, for most cases, the inclusion of the local CD term substantially improves the success probability. Also, we consider the hybrid classical-quantum algorithms for tackling hard instances of the portfolio optimization problem. In particular, we considered QAOA and DC-QAOA methods and showed that CD-assisted QAOA gives better performance than the naive approach. However, for random initialization, finding optimal parameters for QAOA and DC-QAOA is a challenging task due to the highly non-convex nature of the cost landscape. 

In conclusion, adding the approximate CD terms provides a drastic enhancement for solving the portfolio optimization problem by finite-time adiabatic evolution and also for hybrid classical-quantum algorithms. The present study considers only 2-local CD terms. Extending this to higher-order terms is expected to give further enhancement. As an outlook, to consider the development of a digital-analog quantum computing encoding, on top of the digitized-counterdiabatic quantum algorithm for solving the same problem, may approach us to quantum advantage for industrial use cases in the NISQ era.

\begin{acknowledgments}
The authors acknowledge support from projects STCSM (2019SHZDZX01-ZX04 and 20DZ2290900), SMAMR (2021-40) and Junta de Andaluc\'ia (P20-00617).
\end{acknowledgments}

\end{document}